\documentstyle[twoside,fleqn,espcrc2]{article}
\newcommand{\simNinfty}{\begin{array}{c}\;\\[-5pt]\sim\\[-5pt]
{\scriptstyle N\to\infty}\end{array}}
\newcommand{\be}{\begin{equation}}
\newcommand{\ee}{\end{equation}}
\newcommand{\bea}{\begin{eqnarray}}
\newcommand{\eea}{\end{eqnarray}}
\newcommand{\rme}{{\mathrm e}}

\newcommand{\AmS}{{\protect\the\textfont2
  A\kern-.1667em\lower.5ex\hbox{M}\kern-.125emS}}

\title{Generalized two-dimensional Yang-Mills theory is a matrix
string theory}

\author{M. Bill\`o, M. Caselle, A. D'Adda\address{Dipartimento di Fisica
    Teorica\\ Universit\`a di Torino\\ Via P. Giuria 1, I-10125 Torino, Italy}
    and P. Provero\address{Dipartimento di Scienze e Tecnologie Avanzate\\
Universit\`a del Piemonte Orientale\\I-15100 Alessandria, Italy}}
\begin{document}
\begin{abstract}
We consider two-dimensional Yang-Mills theories on arbitrary
Riemann surfaces. We introduce a  generalized Yang-Mills action,
which coincides with the ordinary one on flat surfaces but differs from it
in its coupling to two-dimensional gravity.
The quantization of this theory in the unitary gauge can be consistently
performed taking into account all the topological sectors arising from
the gauge-fixing procedure. The resulting theory is naturally interpreted
as a Matrix String Theory, that is as a theory of covering maps from a
two-dimensional world-sheet to the target Riemann surface.\\
\vskip 8pt
{\bf Report-no DFTT 2/2000.} Presented by A. D'adda at the 
{\sl Third Meeting on Constrained Dynamics and Quantum Gravity},
Villasimius (Sardinia, Italy) September 13-17, 1999; to appear in the
proceedings.   
\end{abstract}


\maketitle

\section{INTRODUCTION}

In the early seventies 't Hooft \cite{thooft} established that in the limit of
a large number $N$ of colours gauge theories are dominated by planar
diagrams, namely by Feynman diagrams with the topology of a sphere,
and that diagrams of genus $g$ are suppressed by a factor
$[\frac{1}{N}]^{2g -2}$. The expansion in powers of $\frac{1}{N}$
can thus be reinterpreted as a perturbative string expansion, with
string  coupling constant 
$\frac{1}{N}$.
In two space-time dimensions, gauge fields have no
physical degrees of freedom and pure Yang-Mills theory (YM2) is
invariant under area preserving diffeomorphisms, thus revealing its
almost topological nature. As a consequence of these features it can
be exactly solved on arbitrary Riemann surfaces, both in a lattice
\cite{m75,r90} and continuum \cite{w91,w92,bt92} formulation.
The interpretation in terms of a string theory of this solution was
provided in a series of beautiful papers \cite{g93,gt93} where it was
shown that the coefficients of the $\frac{1}{N}$ expansion of the YM2
partition function count the number of maps (without foldings) from
a two-dimensional world-sheet of  genus-$g$ to the Euclidean space-time,
which is a Riemann surface of genus $G$.
In two dimensions the number of maps is the number of string
configurations and, as each map is weighed by the area of its world
sheet, one can conclude that in this limit YM2 is a two-dimensional
string theory.
\par
A string theory can be associated to a (generalized)  YM2 theory in a completely
different way, inspired by Matrix String Theory (MST) \cite{m97,dvv97}.
This new approach, introduced in Refs. \cite{bcdp99,bcdp99b} for the case
of the cylinder and the torus, and generalized in \cite{bdp99} to arbitrary
Riemann surfaces, will be the subject of this talk.
It originates quite naturally from quantization of YM2 in a non-propagating
gauge, the unitary gauge, but it requires a generalization of YM2 in the 
sense of \cite{dls94,gsy95}. On the other hand, the string interpretation is
not related in this case to the expansion in powers of $\frac{1}{N}$,
and it is valid for any value of $N$.
\par
The plan of this contribution is the following. In Sec. 2 we
review the exact solution of lattice YM2, and its interpretation
as a string theory in the large-$N$ limit.
Sec. 3 contains a general discussion of MST; Sec. 4 discusses the
quantization of YM2 in the unitary gauge. In Sec. 5 we introduce the
generalized YM2 and its interpretation as a MST. In Sec. 6 we extend the
previous considerations to Riemann surfaces with boundaries, while in
Sec. 7 we introduce a lattice gauge theory whose partition function
coincides with the one of our generalized YM2. Sec. 8 is devoted to
some concluding remarks and outlines of future research.

\section{YM2 ON ARBITRARY RIEMANN SURFACES}
The exact expression of the partition function of YM2 on a general
Riemann surface was first obtained on the lattice by Rusakov
\cite{r90}, based on previous work by Migdal \cite{m75}. It is
convenient to consider the lattice theory defined by the heat-kernel
action: if the product of the group elements
around a plaquette is $V$, its Boltzmann weight is
\be
\label{uno}
\rme^{-S_{t}(V)}=\sum_{R}d_{R}\chi_{R}(V)\rme^{-\frac{t}{2}C_{R}}~,
\ee
where the sum is over all the irreducible representations $R$ of the
gauge group, $d_{R}$ is the dimensionality and $C_{R}$ the quadratic
Casimir of the representation, $\chi_{R}(V)$ the character of $V$, and
$t=g_{\mathrm{YM}}^{2}A$, with $g_{\mathrm{YM}}$ the gauge coupling and $A$ the area of the
plaquette. Notice that $t$ is a-dimensional and it measures the area in
units of the dimensional coupling constant $g_{\mathrm{YM}}$.
\par
This action has a unique property: if we integrate over a link
variable $U$ belonging to two adjoining plaquettes $A_1$ and $A_2$ we
obtain exactly the Boltzmann weight corresponding to the heat kernel
action for the resulting plaquette $A_1 \cup A_2$.
In formulae this reads:
\be
\int dU \rme^{-S_{t_{1}}(V_{1}U^{\dagger})-S_{t_{2}}(UV_{2})}=
\rme^{S_{t_{1}+t_{2}}(V_{1}V_{2})}~.
\label{renorm}
\ee
This property is a simple consequence of the orthogonality of the
characters, and it allows to reduce, by successive integrations,
any kernel with an arbitrary number of boundaries to a single plaquette
with the sides suitably identified, and eventually to compute it
exactly by using again well known properties of the characters.
On the other hand, the continuum limit can be reached smoothly by adding
more and more links to make the lattice spacing smaller and smaller, and the
same renormalization invariance (\ref{renorm}) insures that the result obtained
on the lattice coincides with the one in the continuum.
\par
The result of the calculation for a general Riemann surface of genus $G$ with an
arbitrary number $p$  of boundaries  is 
\bea
&&K_{G,p}(g_{1},\dots,g_{p};t)\nonumber\\
\ \ &&=\sum_{R}d_{R}^{2-2G-p}\chi_{R}(g_{1})
\dots\chi_{R}(g_{p})\rme^{-\frac{t}{2}C_{R}}~,
\eea
where the $g_{i}$'s are the group elements defined on the boundaries.
\par
For $p=0$ we obtain the partition function
\be
Z_{G}(t)=\sum_{R}d_{R}^{2-2G}\rme^{-\frac{t}{2}C_{R}}~.
\ee
{}From now on we will consider the gauge group $\mathrm{U}(N)$ only, whose
irreducible representations are labeled by $N$ integers
\be
n_{1}>n_{2}>\dots>n_{N}~.
\ee
The corresponding dimensions $d_R $ and Casimir invariants $C_R$
are given by:
\bea
d_{R}&=&\prod_{i>j}(n_{i}-n_{j})~,\\
C_{R}&=&\sum_{i}n_{i}^{2}~,
\eea
so that the partition function reads
\bea
Z_{G}(N,t)&=&\sum_{n_{1}>n_{2}>\dots>n_{N}}
\prod_{i>j}(n_{i}-n_{j})^{2-2G}\nonumber\\
&\cdot&\rme^{-\frac{t}{2}\sum_{i}n_{i}^{2}}~.
\label{pf}
\eea
In Refs. \cite{g93,gt93} the large-$N$ limit of Eq. (\ref{pf}) was
investigated. The limit was taken, as introduced by `t Hooft \cite{thooft},
by sending $N\to\infty$ and $t\to 0$ while keeping $\tilde{t}\equiv Nt$
constant.
\par
The result of the analysis is that the partition function (\ref{pf}) in
such limit can be interpreted as the theory of a string living in the
same two-dimensional target space $M_{G}$ as the original gauge theory. Each
string configuration is described by a covering map from a
two-dimensional world-sheet $W_{g}$ into $M_{G}$.
The admissible string configurations have no folds, so that the area
of the world-sheet is an integer multiple of the area of the target
manifold. The covering maps are allowed to have branch points.
\par
The equivalence of YM2 with the  string theory described above is expressed
by the following equation\footnote{We consider here a simplified case where
the sum over representations  is restricted to ``chiral'' representations of
$\mathrm{U}(N)$. Also the so called $\Omega$-point type of singularities, 
which are present for $G>1$, have been neglected. See \cite{gt93} for details.}:
\be
Z_{G}(N,t)\simNinfty
\sum_{i,g,n}N^{2-2g}\rme^{-n\tilde{t}/2} (\tilde{t})^{i} \omega^{n,i}_{g,G}~.
\label{gros}
\ee
In (\ref{gros}),  $\omega^{n,i}_{g,G}$ is the number of $n$-coverings of the
target manifold with genus $g$ and $i$ branch points.

\section{MATRIX STRING THEORIES (MST)}
The interpretation  of a generalized YM2 as a string theory that we want
to present is inspired to the mechanism through which string states emerge
in MST. We begin by briefly introducing a definition of 
Matrix String Theories sufficiently general to encompass both the original 
work \cite{dvv97}
and our description of YM2.
Consider a gauge theory defined on a Riemann surface, and
containing at least one
field $F$ transforming in the adjoint representation of the gauge group:
in the case of $\mathrm{U}(N)$, $F$ is a hermitean $N\times N$ matrix. A
possible gauge choice consists in diagonalizing $F$, that is in
conjugating it into an element of the Cartan sub-algebra.
\par
This gauge choice is characterized by a Gribov ambiguity related to
the Weyl group, namely the group of invariance of the Cartan
sub-algebra, which in our case is the symmetric group $S_N$:
in fact given a hermitian matrix $F$ there exist $N!$ matrices that
diagonalize $F$, corresponding to all possible permutations of the
eigenvalues.
\par
So in order to fix the gauge completely one has to choose in each point
one of the $N!$ Gribov copies. It is easy to see that in general  this cannot
be done smoothly on the whole Riemann surface whenever the first homotopy
group of the surface is non-trivial.
\par
In fact let $\gamma$ be a homotopically non-trivial closed path based
in a point $x$, and choose one of the Gribov copies in
$x$, say the one, named {\em standard} in the following,
where $F= {\rm diag}(\lambda_i)$ with
$\lambda_{1}(x)\ge\lambda_{2}(x)\ge\dots\ge\lambda_{N}(x)$. If we
diagonalize $F$ smoothly along $\gamma$ and we return to $x$, we will
in general end up on a different Gribov copy, related to the standard
one by a permutation $P(\gamma)$.
\par
In this way we can associate a permutation to each homotopy class of
closed paths based in $x$. It is easy to convince oneself that this
defines a group homomorphism
\be
P:\pi_{1}(M;x)\to S_{N}~.
\label{homo}
\ee
We conclude that each configuration of the adjoint field $F$ belongs
to a {\em topological sector} identified by a group homomorphism
(\ref{homo}).
\par
Consider for example the case of the torus: the homotopy group is
generated by two cycles $a$ and $b$ with the relation
\be
ab=ba~.
\ee
Therefore to each configuration of the field $F$ and to each base
point $x$ we can associate a topological sector defined by a pair
of {\em commuting} permutations $(P_{1},P_{2})$; changing the base
point $x$ results in changing the pair  $(P_{1},P_{2})$ into a
conjugate pair $(QP_{1}Q^{-1}, QP_{2}Q^{-1})$.
\par
The homomorphisms (\ref{homo}) are in one-to-one correspondence with
the $N$-coverings of the surface. This correspondence can be easily
visualized by imagining each eigenvalue of $F$ as living on one of
the $N$ sheets of the covering. Therefore we seem to be in a position
to conclude that for a general Riemann surface, any theory with a
$\mathrm{U}(N)$ gauge symmetry and at least one field transforming in the
adjoint representation can be written as a sum over coverings of the
surface, that is as a string theory.
\par
While this is true in principle, it is of real
interest only if the resulting string theory is free or weakly
coupled, that is if the different sheets of the $N$-covering ({\em
i.e.} the different eigenvalues of $F$) do not
interact among themselves or interact weakly.
\par
In general, however, strong interactions between the eigenvalues of
$F$ do occur, coming from two different sources:
\begin{enumerate}
\item from the functional integration over the other fields of the
theory;
\item from the Faddeev-Popov determinant induced by the gauge-fixing.
\end{enumerate}
The latter source of interactions is obviously always present, since
the Faddeev-Popov determinant is just the Vandermonde determinant of
the eigenvalues:
\be
\Delta_{\mathrm{FP}}=
\exp\int d^{2}x\; 2\sqrt{g}  \sum_{i>j}\log|\lambda_{i}-\lambda_{j}|~.
\ee
\par
Therefore, for the gauge theory to be a (weakly coupled) string theory
{\em the Vandermonde determinant must be canceled by the integration
over other fields}.
\par
In Ref. \cite{m97,dvv97}, a MST
was introduced as a non-perturbative formulation of
type-IIA superstring theory. It is described in terms of a
supersymmetric, two-dimensional $\mathrm{U}(N)$ gauge theory with action
\bea
S&=&\int d^{2}\xi {\rm Tr} \left[ D_{\mu}X^{i}D^{\mu}X^{i}+\frac{1}{2}
g_{\mathrm{s}}^{2}F_{\mu\nu}F^{\mu\nu}\right.\nonumber\\
&+&\frac{1}{2g_{\mathrm{s}}^{2}}\left[X^{i},X^{j}\right]^{2}-i\bar{\theta}
\gamma^{\mu}D_{\mu}\theta\nonumber\\
&-&\left.\frac{1}{g_{\mathrm{s}}}\theta^{T}\Gamma_{i}
\left[X^{i},\theta\right]\right]~,
\eea
where $g_{\mathrm{s}}$ is the string coupling, $X^{i}$ $(i=1,\dots,8)$ are
$N\times N$ hermitean matrices representing the transverse space-time
coordinates, $\theta$ their fermionic superpartners, and $F$ is the
$\mathrm{U}(N)$ field strength.
\par
In the limit $g_{\mathrm{s}}\to 0$  all the fields $X$ and $\theta$ commute
and can be diagonalized simultaneously. This diagonalization produces
the topological sectors as described above. The
cancellation of the Vandermonde determinant is achieved through
supersymmetry: the contribution of the fermionic fields $\theta$
cancels the one of the $X$ fields, so that the strong interaction
between the eigenvalues disappears and the theory indeed describes
free strings propagating in 8 transverse dimensions. For nonzero
$g_{\mathrm{s}}$, weak string interactions of order $g_{\mathrm{s}}^{2}$ are 
generated by the non-diagonal terms.

\section{YM2 IN THE UNITARY GAUGE}
In this section we will discuss the quantization of YM2 in the unitary gauge,
namely in the gauge where the field strength $F$ is diagonal. This condition
does not fix the gauge completely as it leaves a ${\mathrm{U}}(1)^N$ gauge invariance,
so that further gauge fixing conditions for the abelian part have to be
introduced separately.  The unitary gauge is a non-propagating gauge (the
non-diagonal components of the ghost-antighost fields do not propagate) and
in four dimensions it makes the theory non-manifestly renormalizable.
As we shall see, also in two dimensions this gauge is plagued by divergences
that occur when two eigenvalues coincide. However, these can be eliminated by
considering a suitable generalized YM2 (in  the sense of \cite{dls94,gsy95}).
This procedure also eliminates the interaction between different eigenvalues
induced by the Vandermonde determinants, thus ensuring that the strings
associated to the topological sectors discussed in the previous section are
essentially free.
\par
The starting point is the first order formalism for YM2 on a genus-$G$ Riemann
surface: the partition function is
\bea
Z&=&\int [dA][dF] \exp\left[-\int_{\Sigma_G}d\mu V(F)\right.\nonumber\\
&+&\left.{\mathrm{i}}\,{\rm Tr}\int f(A) F\right]~,
\eea
where
\be
f(A)=dA+A\wedge A~,
\ee
F is an auxiliary field transforming in the adjoint representation of 
$\mathrm{U}(N)$,
and $V(F)$ is a gauge-invariant potential, depending on the eigenvalues of
$F$ only. This defines a {\em generalized} Yang-Mills theory
\cite{dls94,gsy95};
ordinary YM2 is recovered by choosing
\be
V(F)={\rm Tr} \frac{t}{2} F^2
\ee
and integrating over $F$ to obtain the usual second-order formalism.
\par
To determine whether such a theory is a MST, we have to
\begin{enumerate}
\item{fix the gauge where $F$ is diagonal;}
\item{check whether the Vandermonde determinant coming from the gauge
fixing can be canceled by terms coming from other fields.}
\end{enumerate}
\par
The BRST-invariant action for the unitary gauge is of the form
\be
S_{\mathrm{BRST}}=S_{\mathrm{Cartan}}+S_{\mathrm{off-diag}}~,
\ee
where $S_{\mathrm{Cartan}}$ depends only on the eigenvalues of 
$F$ and the diagonal components of the $A$ fields:
\be
S_{\mathrm{Cartan}}=
\int_{\Sigma_G}\left[V(\lambda)-{\mathrm{i}}\sum_{i=1}^{N}\lambda_i dA^{(i)}~,
\right]
\ee
while $S_{\mathrm{off-diag}}$ is given by
\bea
S_{\mathrm{off-diag}}&=&
\int d\mu \sum_{i>j}(\lambda_i-\lambda_j)\Bigl[\frac{1}{2}
\epsilon^{ab}\hat{A}^{ij}_a\hat{A}^{ji}_b\nonumber\\
&+&{\mathrm{i}}\,c^{ij}\bar{c}^{ji}+
{\mathrm{i}}\,c^{ji}\bar{c}^{ij}\Bigr]~,
\eea
where
\be
\hat{A}^{ij}_a\equiv E^\mu_aA^{ij}_\mu~,
\ee
with $E_a^\mu$ the inverse vierbein; $c$ and $\bar{c}$ are the ghost and
anti-ghost fields. If YM2 is a MST, the integrations over
$\hat{A}$ and $c,\bar{c}$ must cancel each other.
\par
Note that at the classical level  $S_{\mathrm{off-diag}}$ is invariant under the
following transformation, depending on four Grassmann-valued parameters
$\eta$, $\zeta$, $\xi$, $\chi$:
\bea
\delta\hat{A}_0&=&{\mathrm{i}}(\eta c+\zeta \bar{c})~,\\
\delta\hat{A}_1&=&{\mathrm{i}}(\xi c+\chi \bar{c})~,\\
\delta c&=& -\chi\hat{A}_0+\zeta \hat{A}_1~,\\
\delta\bar{c}&=&-\xi \hat{A}_0+\eta \hat{A}_1~.
\eea
This ``supersymmetry'' does indeed seem to guarantee the cancellation of the
contribution of $S_{\mathrm{off-diag}}$.
\par
However, it is known \cite{bt94}
that this supersymmetry is in general broken by
quantum effects, the anomaly being proportional to the scalar curvature
of $\Sigma_G$:
\bea
&&\int [dc][d\bar{c}][dA]\rme^{-S_{\mathrm{off-diag}}}\nonumber\\
&&\ \ =\exp\frac{1}{8\pi}\int_{\Sigma_G}
d\mu R \sum_{i>j} \log|\lambda_i-\lambda_j|~.
\label{anomaly}
\eea
This anomaly is essentially due to the fact that for a curved surface
(twice) the number of 0-forms (the $c$ and $\bar{c}$ fields) does not
equal the number of 1-forms (the field $A$).
\par
Inserting this result into the partition function we obtain an effective
theory for the diagonal degrees of freedom that exhibit a residual 
$\mathrm{U}(1)^N$ gauge invariance:
\bea
Z&=&\int [dA^{(i)}][d\lambda_i]\exp -\int_{\Sigma_G}\Bigl[\tilde{V}
(\lambda)d\mu
\nonumber\\
&-&{\mathrm{i}}\sum_{i}\lambda_i dA^{(i)}\Bigr]~,
\label{pfug}
\eea
with
\be
\tilde{V}(\lambda)=V(\lambda)-\frac{1}{8\pi}\int_{\Sigma_G}R\sum_{i>j}
\log|\lambda_i-\lambda_j|~.
\label{potential}
\ee
The last term at the r.h.s. of (\ref{potential}) is singular when two
eigenvalues coincide and the problem of a consistent regularization
leading to the standard expression (\ref{pf})  for the partition function
is essentially unsolved.
\par
According to  Ref. \cite{bt94}, the procedure to recover  (\ref{pf})
from (\ref{pfug}) involves two steps:
\begin{enumerate}
\item{The ${\mathrm{U}}(1)^N$ gauge fixing and the functional 
integration over the
$A^{(i)}$ fields are performed, and the result is equivalent to replacing
all the eigenvalues $\lambda_i(x)$ with integer-valued constants $n_i$
(the flux of the abelian field strength) and summing over the
$n_i$'s. One obtains
\be
Z=\sum_{n_1,\dots,n_N}\prod_{i>j}\left(n_i-n_j\right)^{2-2G}\rme^{-V(n)}~.
\label{pfint}
\ee}
\item{The terms with two integers $n_i$ coinciding, to be referred to in the
following as {\em non-regular terms}, are divergent and are essentially
discarded by hand.}
\end{enumerate}
In this way, with $V(n)=\frac{t}{2}\sum_i n_i^2$ one recovers
Eq. (\ref{pf}).
\par
Discarding non-regular terms also means discarding the non-trivial topological
sectors described in Sec. 3. In fact in a non-trivial topological sector there
is at least one homotopically non-trivial closed path that induces a non-trivial
permutation on the eigenvalues. As the eigenvalues are constant integers, this
implies that at least two eigenvalues coincide.
In the following section we propose an alternative approach, that allows to
eliminate all divergences while keeping all non-trivial sectors at the price,
however, of modifying the coupling of YM2 to gravity.
\section{GENERALIZED YM2 IS A MATRIX STRING THEORY}
The divergences of the theory in the unitary gauge can be eliminated, rather
than by the {\it ad hoc} procedure of Ref. \cite{bt94}, by a redefinition of the
theory, namely by using the arbitrariness of the potential $V(\lambda)$ to
absorb the divergent term due to the anomaly. 
In other words since all choices of $V(\lambda)$ define a legitimate
generalized Yang-Mills theory, it is apparent that there is a specific choice
in which the anomaly cancels and one obtains a Matrix String Theory: namely we
can choose $V(\lambda)$ in such a way that the anomaly term cancels and
$\tilde{V}(\lambda)$ is regular when two eigenvalues coincide.
\par
For example we can choose $V(\lambda)$ such that
\be
\tilde{V}(\lambda)=\frac{t}{2}\sum_i\lambda_i^2~.
\label{vtil}
\ee
With this choice we define a theory whose action coincides with the one of 
ordinary YM2 on flat surfaces, but differs from it when coupled to 
two-dimensional gravity (recall that the anomaly (\ref{anomaly}) is proportional
to the curvature of the Riemann surface).
\par
The partition function of the generalized YM2 defined by the potential
(\ref{vtil}) consists of topological sectors which are in one-to-one 
correspondence with  the inequivalent coverings of the target manifold.
There is no  interaction among different eigenvalues, so the partition function
describes a theory of non interacting strings.
It is clear from Eq. (\ref{pfug}) that a ${\mathrm{U}}(1)$ gauge theory is 
defined on each  connected component of the covering, 
that is of the world sheet. 
The product of the partition functions of these ${\mathrm{U}}(1)$ 
gauge theories gives the Boltzmann weight of the covering. 
Although a compact expression for the partition function is in general not
available we present here the recipe for writing it down for a general Riemann
surface without boundaries (the case of a surface with boundaries will be
discussed in Sec. 6). It is convenient to work directly on the generating
function of the partition functions for arbitrary $N$, namely the  
grand-canonical partition function. This is defined as
\be
Z(G,q)=\sum_{N=0}^{\infty}Z_N(G)q^N~,
\ee
where $Z_N(G)$ is the $\mathrm{U}(N)$ partition function. 
For a given Riemann surface
$\Sigma_G$ of genus $G$ one proceeds as follows:
\begin{enumerate}
\item
Write the grand-canonical partition function of unbranched coverings
of $\Sigma_G$:
\be
Z^{(\mathrm{cov})}(G,q)=\sum_{k=0}^{\infty} Z_k^{(\mathrm{cov})}(G) q^k~,
\ee
where $Z_k^{(\mathrm{cov})}(G)$ is the number of inequivalent $k$-coverings of
$\Sigma_G$. $Z^{(\mathrm{cov})}(G,q)$ has been studied in the literature
\cite{gt93,ks97,ksw98} and is known in closed form.
\item
Take the logarithm of  $Z^{(\mathrm{cov})}(G,q)$ 
to obtain the free energy
\bea
F^{(\mathrm{cov})}(G,q)&=&\log Z^{(\mathrm{cov})}(G,q)\nonumber\\
&=&\sum_{k=0}^{\infty} F_k^{(\mathrm{cov})}(G) q^k~,
\eea
where $F_k^{(\mathrm{cov})}(G)$ is the number of {\em connected} $k$-coverings of
$\Sigma_G$. No closed expression is known for $F^{(\mathrm{cov})}(G,q)$.
\item
Associate to each connected covering the partition function of the
${\mathrm{U}}(1)$ gauge theory living on it, to obtain the free energy of the
generalized YM2:
\be
F(G,q,t)=\sum_{k=0}^{\infty} F_k^{(\mathrm{cov})}(G) 
Z_{{\mathrm{U}}(1)}(k,t) q^k~.
\ee
Here $Z_{{\mathrm{U}}(1)}(k,t)$ is the partition function of a (generalized)
${\mathrm{U}}(1)$ gauge theory on a Riemann surface of area $k t$ 
(in units of the gauge coupling constant). 
$Z_{{\mathrm{U}}(1)}(k,t)$ does not depend on the genus and for the
quadratic potential (\ref{vtil}) it reads\footnote{The generalization to
arbitrary potentials is straightforward}:
\be
Z_{{\mathrm{U}}(1)}(k,t)=\sum_{n=-\infty}^{\infty}\rme^{-\frac{1}{2} kt n^2}~.
\ee
Therefore
\be
F(G,q,t)=\sum_{n=-\infty}^{\infty}F^{(\mathrm{cov})}
(G,\rme^{-\frac{1}{2}t n^2}q)~.
\ee
\item
The gauge theory partition function is then
\bea
Z(G,q,t)&=&\rme^{F(G,q,t)}=\sum_{N=0}^{\infty}Z_N(t,G)q^N\nonumber\\
&=&\prod_n Z^{(\mathrm{cov})}(G,\rme^{-\frac{1}{2}t n^2}q)~.
\eea
\end{enumerate}
\par
As an example, we will treat explicitly the case of the torus. This is the
only case (except for the trivial one of the sphere) where a closed expression
is available for the number of connected coverings.
As discussed in Sec. 3,  the coverings of the torus are in one-to-one
correspondence with the pairs of commuting permutations. The number of
such pairs is $N!p(N)$ where $p(N)$ is the number of partitions of $N$.
Therefore there are $p(N)$ inequivalent coverings (that is, not counting
as different two coverings that can be obtained from each other by simple
relabeling of the sheets) and we have:
\bea
Z^{(\mathrm{cov})}(G=1,q)&=&\sum_{N=0}^{\infty} p(N) q^N\nonumber\\
&=&\prod_{k=1}^{\infty}\frac{1}{1-q^k}~.
\eea
For the connected coverings we have
\be
F^{(\mathrm{cov})}(G=1,q)=\sum_{N=1}^{\infty}\sum_{r|N}\frac{1}{r} q^N~,
\ee
where the second sum is extended to all divisors of $N$ (including $N$).
\par
Therefore we obtain for the YM2 free energy the expression 
\be
F(G=1,q,t)=\sum_{n=-\infty}^\infty \sum_{N=1}^\infty
\sum_{r|N}\frac{1}{r}\rme^{-\frac{1}{2} N t n^2}q^N
\ee
and for the partition function
\be
Z(G=1,q,t)=\prod_{n=-\infty}^\infty\prod_{k=1}^{\infty} 
\frac{1}{1-q^k\rme^{-\frac{1}{2} k t n^2}}~.
\ee
This result does not coincide with the expression (\ref{pfint}) of the
partition function for $G=1$. To recover the ``usual'' partition function
on the torus, one has to consider only the topological sectors corresponding
to pairs of commuting permutations of the special form $(1,P)$, weighted
with a factor $(-1)^{n_c}$ where $n_c$ is the number of connected components
of the covering. One can check that this prescription effectively cancels all
non-regular terms from Eq. (\ref{pfint}).
\par
Therefore for the case of the torus, where the Riemann curvature vanishes and
our generalized YM2 coincides with the usual one at the classical level, the 
inclusion of all the topological sectors leads to a different quantum theory 
which is richer in structure and  closely related to the MST of Ref. 
\cite{dvv97}, as shown in Refs. \cite{kv98,gs99}.
\section{SURFACES WITH BOUNDARIES}
Consider now a Riemann surface $\Sigma_{G,p}$ of genus $G$ with $p$ boundaries.
The arguments of the preceding sections hold essentially unaltered, and the 
generalized YM2 can be written as a theory of $N$-coverings of
$\Sigma_{G,p}$ with a ${\mathrm{U}}(1)$ gauge theory living on each 
connected component of the covering.
\par
The natural quantity to be studied is the kernel that will depend on $p$
states defined on each boundary. For any given boundary, a state is completely
specified by assigning:
\begin{enumerate}
\item{A {\em conjugacy class} of permutations, defined as the conjugacy class
of the permutation $P(\gamma)$ where $\gamma$ is any closed curve that can
be continuously deformed into the boundary. It is given in terms of $N$
integers $r_l$ $(l=1,\dots,N)$ satisfying $\sum_l lr_l=N$ where $r_l$ is the 
number of cycles of length $l$ in $P(\gamma)$.}
\item{A ${\mathrm{U}}(1)$ holonomy associated to each {\em cycle} of the 
conjugacy class defined above.}
\end{enumerate}
The procedure to construct the kernels follows closely the one described
above for surfaces without boundaries: one constructs the free energy of
the coverings of $\Sigma_{G,p}$ and associates to each connected $k$-covering
a ${\mathrm{U}}(1)$ kernel
\be
K_{{\mathrm{U}}(1)}(kt,\phi)=\sum_{n=-\infty}^{\infty}
\rme^{-\frac{1}{2} k t n^2+{\mathrm{i}}n\phi}~.
\ee
Exponentiation then produces the generalized YM2 kernel. 
Details can be found in Ref. \cite{bdp99}.
One must check the consistency of the theory, namely that  the set of kernels
constructed in this way produces consistent results when the surfaces are cut 
and sewn to produce new ones. This property is obviously true for the pure 
theory of coverings and it can be proved that the introduction of the 
${\mathrm{U}}(1)$
gauge  theory on each connected world sheet does not spoil it.

\section{LATTICE EFFECTIVE THEORY}
We have seen in the previous sections that, after integration over the
non-diagonal components of the fields, our generalized YM2 theory is 
described by an effective theory of coverings with a ${\mathrm{U}}(1)$ 
gauge theory on each connected world sheet.
In this section we show that this effective theory is equivalent to a lattice 
theory whose gauge group is the semi-direct product of $S_N$ and 
${\mathrm{U}}(1)^N$.
Consider first a surface with the topology of a disc. Since its fundamental
group is trivial, the only $N$-covering consists of $N$ disjoint copies of
the disc: all closed paths are mapped into the identity by the homomorphism
(\ref{homo}). From the discussion of the previous section it follows that the
state on the boundary is given in terms of $N$ ${\mathrm{U}}(1)$ invariant 
angles $\phi_i$. The kernel is then
\be
K((P,\phi),t)=\delta(P)\sum_{n_i}\rme^{\sum_{i=1}^N
\left({\mathrm{i}}n_i\phi_i-tv(n_i)\right)}~.
\label{disc}
\ee
\par
Consider now the subgroup ${\cal G}_N$ of $\mathrm{U}(N)$ given by the 
matrices of the form
\be
(P,\phi)={\rm diag}\left(\rme^{{\mathrm{i}}\phi_1},
\dots,\rme^{{\mathrm{i}}\phi_N}\right)\;P~,
\ee
where the $\phi_i$'s are real and $P$ is a permutation matrix:
\be
P_{ij}=\delta_{iP(j)}~.
\ee
The group product reads
\be
(P,\phi)(Q,\theta)=(PQ,\phi+P\theta)~,
\ee
which shows that ${\cal G}_N$ is a semi-direct product of $S_N$ and 
${\mathrm{U}}(1)^N$.
\par
The kernel (\ref{disc}) is invariant under gauge transformations 
belonging to ${\cal G}_N$:
\be
K\left((Q,\theta)(P,\phi)(Q,\theta)^{-1}, t\right) =
K\left((P,\phi), t\right)~.
\label{gauge}
\ee
Moreover it shares with the heat-kernel action (\ref{uno}) 
of lattice gauge theory the 
invariance under renormalization group transformations:
\bea
&&\sum_{Q\in S_N}\int \prod_i \frac{d\theta_i}{2\pi} K\left((P_1,\phi_1)
(Q,\theta),t_1\right)\nonumber\\
&&\hskip 1.5cm\cdot K\left((Q,\theta)^{-1}(P_2,\phi_2), t_2
\right)\nonumber\\
&&\;\; =K\left((P_1, \phi_1)(P_2, \phi_2), t_1+t_2\right)~.
\label{renorm2}
\eea
\par
The kernel (\ref{disc}) can thus be used as a heath-kernel type of action
to construct a lattice gauge theory for the group $\mathcal{G}_N$, with
partition function  
\begin{equation}
Z_{\rm latt} = \prod_{\alpha}\sum_{P_{\alpha}} \int_0^{2 \pi} d\phi_{\alpha}
\prod_{\rm pl} K\left((P_{\rm pl},\phi_{\rm pl}), t
\right)~,
\label{lattpart}
\end{equation}
where $\alpha$ enumerates the links in the lattice, and
the product of all the elements $(P_\alpha,\phi_\alpha)$ of ${\cal G}_N$ 
along a plaquette defines 
the plaquette variable $(P_{\rm pl},\phi_{\rm pl})$. 
In complete analogy with the solution of YM2
obtained in~\cite{r90}, one can solve exactly
the theory by reducing the surface to a single plaquette with suitably
identified links. As shown in \cite{bdp99}, the resulting expression coincides
with the one that is obtained, via the steps discussed in the
previous Section, by including all the topological sectors in the 
MST corresponding to our generalized YM2 theory.
\section{CONCLUSIONS AND FURTHER DEVELOPMENTS}
Quantization of YM2 in a non-propagating gauge, such as the unitary gauge,
leads to non-manifestly renormalizable divergences as well as to the appearance 
of a structure of topological sectors, associated to the $N$  coverings of the
target manifold ; this strongly suggests the presence of an
underlying string theory.
In order to recover the standard partition function obtained with other
gauge choices, the divergences have to be removed more or less by hand, thus
suppressing the topological sectors even in the case of torus where they are not
related to any divergence.
\par 
However we found that the divergences can be removed, and the topological
sectors maintained, by considering a generalized theory whose action coincides
with standard one for vanishing Riemann curvature. Indeed one can say that the
generalized theory differs from the conventional one for a non-standard coupling
with gravity. In this approach the string theory 
does not emerge from the large $N$
limit, but directly for any value of $N$ from the topological structure of the
theory as in MST. It is not a chance that for the case of the torus our 
partition function is closely related to the one of the full MST.
\par
These considerations suggest some lines along which future research might
develop. Two of them seem the most natural ones:
\begin{enumerate}
 \item{The string theory considered so far is a theory of free strings. It is
natural to introduce interactions by including branched coverings, that is the 
possibility for strings to split and join. A branch point can be seen as a 
boundary where all the ${\mathrm{U}}(1)$ invariant angles are set to zero. 
In this sense our treatment of surfaces with boundaries includes branch 
points as well. 
However, the most interesting case is the one in which a {\em continuum} 
distribution of branch points is introduced, as was done in Refs. 
\cite{ks97,ksw98} for the pure theory of coverings.}
\item{Other non-propagating gauges would require a different tuning of the
potential to eliminate the divergences and may lead to more general theories.
For instance a gauge choice where, for $N=LM$, the field strength is block
diagonal, with blocks of size $M\times M$, would be equivalent to a theory of
$L$-coverings with a non abelian $U(M)$ gauge theory on the world sheet.
This formulation would include both standard YM2 ($L=1$) and our generalized
theory ($L=N$) as limiting cases.
As a string theory emerges in both cases through completely different
mechanisms, some investigation will be needed to ascertain whether a deeper
relation exists between these two different formulations.}
\end{enumerate}


\begin{thebibliography}{9}
\bibitem{thooft}G. 't Hooft,  Nucl.Phys. { B75} (1974) 461. 
\bibitem{m75} A.A. Migdal, Sov. Phys. JETP { 42} (1975) 413.
\bibitem{r90} B. Ye. Rusakov, Mod. Phys. Lett. { A5} (1990) 693.
\bibitem{w91} E. Witten, Comm. Math. Phys. { 141} (1991) 153.
\bibitem{w92} E. Witten, J. Geom. Phys. { 9} (1992) 303 (hep-th/9204083).
\bibitem{bt92} M. Blau and G. Thompson, Int. J. Mod. Phys. { A7} (1992)
3781.
\bibitem{g93} D. J. Gross, Nucl. Phys. { B400} (1993) 161 (hep-th/9212149).
\bibitem{gt93}  D. J. Gross and W. Taylor, Nucl. Phys. { B400} (1993) 395
(hep-th/9301068).
\bibitem{m97} L. Motl, hep-th/9701025.
\bibitem{dvv97}  R. Dijkgraaf, E. Verlinde and H. Verlinde, Nucl. Phys.
{ B500} (1997) 43 (hep-th/9703030).
\bibitem{bcdp99} M. Bill\'o, M. Caselle, A. D'Adda and P. Provero,
Nucl. Phys. { B543} (1999) 141 (hep-th/9809095).
\bibitem{bcdp99b} M. Bill\'o, M. Caselle, A. D'Adda and P. Provero,
Proceedings Corf\'u Conference, A. Ceresole {\em et al.}, eds.,
Springer LNP 525 (1998) (hep-th/9901053).
\bibitem{bdp99}  M. Bill\'o, A. D'Adda and P. Provero, hep-th/9911249.
\bibitem{dls94} M. R. Douglas, K. Li and M. Staudacher, Nucl. Phys. 
{B420} (1994) 118 (hep-th/9401062).
\bibitem{gsy95} O. Ganor, J. Sonnenschein and S. Yankielowicz,
Nucl. Phys. { B434} (1995) 139 (hep-th/9407114).
\bibitem{bt94}  M. Blau and G. Thompson, in {\em Proc. 1993 Summer School
in High Energy Physics and Cosmology}, E. Gava {\em et al.}, eds.,
World Scientific (1994) (hep-th/9310144).
\bibitem{ks97} I. K. Kostov and M. Staudacher,
Phys.Lett. { B394} (1997) 75 (hep-th/9611011).
\bibitem{ksw98} I. K. Kostov, M. Staudacher and T. Wynter,
Commun. Math. Phys. { 191} (1998) 283 (hep-th/9703189).
\bibitem{kv98} I. Kostov and P. Vanhove, Phys. Lett. { B444} (1998) 196
(hep-th/9809130).
 \bibitem{gs99} G. Grignani and G. W. Semenoff, hep-th/9903246.
\end{thebibliography}
\end{document}